\def\year{2019}\relax
\documentclass[letterpaper]{article} 
\usepackage{aaai19}  
\usepackage{times}  
\usepackage{helvet} 
\usepackage{courier}  
\usepackage[hyphens]{url}  
\usepackage{graphicx} 
\usepackage{graphicx}  
\nocopyright

\usepackage[ampersand]{easylist}

\title{CrowdHub: Extending crowdsourcing platforms for the controlled evaluation of tasks designs}

\author{
Jorge Ram\'irez,\textsuperscript{\rm 1} Simone Degiacomi, \textsuperscript{\rm 1} Davide Zanella, \textsuperscript{\rm 1} \\
\Large \textbf{Marcos Baez,} \textsuperscript{\rm 1} \Large \textbf{Fabio Casati,} \textsuperscript{\rm 1} 
\Large \textbf{Boualem Benatallah} \textsuperscript{\rm 2}\\
\textsuperscript{\rm 1}University of Trento,
\textsuperscript{\rm 2}UNSW Sydney
}

\begin{document}

\maketitle

\begin{abstract}

We present CrowdHub, a tool for running systematic evaluations of task designs on top of crowdsourcing platforms. The goal is to support the evaluation process, avoiding potential experimental biases that, according to our empirical studies, can amount to 38\% loss in the utility of the collected dataset in uncontrolled settings. Using CrowdHub, researchers can map their experimental design and automate the complex process of managing task execution over time while controlling for returning workers and crowd demographics, thus reducing bias, increasing utility of collected data, and making more efficient use of a limited pool of subjects.

\end{abstract}

\section{Background \& Motivation}

A crucial aspect in running a successful crowdsourcing project is identifying an appropriate task design \cite{DBLP:journals/pvldb/JainSPW17}. 
Studies have shown that worker behavior can be influenced by different factors such as task design and presentation \cite{DBLP:conf/chi/SampathRI14}, allocated time to completion
\cite{maddalena2016crowdsourcing}, pricing, and reward schemes
\cite{difallah2014scaling}, human aspects in collaboration and individual biases \cite{drapeau2016microtalk,eickhoff2018cognitive}, and even characteristics of the crowd marketplace and work environment \cite{DBLP:journals/imwut/GadirajuCGD17}. 


However, the potential size of the design space, along with the individual and environmental biases, and the limitations of crowdsourcing platforms, makes it difficult to systematically study tasks designs. As a result, workers still deal with poorly designed tasks \cite{DBLP:conf/ht/GadirajuYB17} that can affect their performance and introduce systematic biases \cite{DBLP:conf/hcomp/FaltingsJPT14}, producing undesirable results for requesters or deterring aggregation models from deriving the right answers \cite{DBLP:conf/hcomp/KamarKH15}.



In this WiP, we explore challenges that arise when evaluating multiple task designs, and we introduce CrowdHub a tool for running crowdsourcing experiments, offering features to overcome these issues.








\section{Challenges in Evaluating Task Designs}



We explored the challenges in evaluating task designs while studying the impact of highlighting support in text classification \cite{ramirez2019}. The goal was to understand \textit{if}, and \textit{under what conditions}, highlighting text excerpts relevant to a given relevance question would improve worker performance. 
This required testing different highlighting conditions (of varying quality) against a baseline without highlighting, given different document sizes and datasets of different characteristics. 
The resulting experimental design featured a combination of \textit{dataset} (3) x \textit{document size} (3) x \textit{highlighting conditions} (6) - a total of 54 configurations. 




Crowdsourcing platforms such as Figure Eight (F8) offer the building blocks to design and run crowdsourcing tasks. In F8, this implies  defining i) data units to classify, ii) gold data to use for quality control, iii) task design, including instructions, data to collect, 
assignment of units to workers,
iv) the target population (country, channels, trust), and v) the cost per worker contribution. 
These features are suitable for running individual tasks, but less so when experimenting with different task designs with a limited pool of workers. 


In order to identify and quantify the experimental bias in running an \textit{uncontrolled} evaluation of task designs, we created individual tasks in F8 for a subset (1 dataset) of the  experimental conditions, and ran them one after another, collecting a total of $6993$ votes from $631$ workers (16 tasks). 
The analysis of the results points to the following issues: 

\begin{figure*}[t]
\centering
\includegraphics[width=\textwidth]{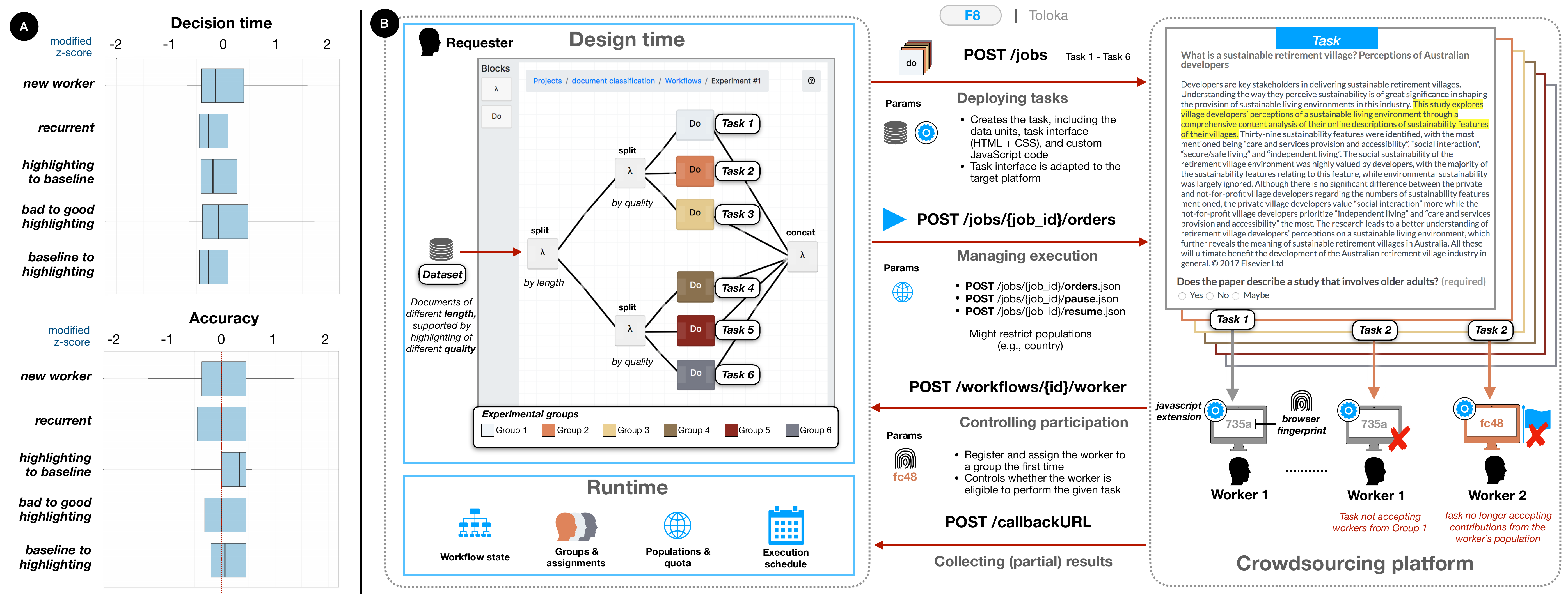}
\caption{ A) Decision time and accuracy for recurrent workers in the highlighting support experiment. Values are normalized to the distribution of \textit{new workers} in each condition. B) Example workflow for a between-subjects design using CrowdHub.}
\label{fig:flow}
\end{figure*}

\begin{easylist}
\ListProperties(Hide=100, Hang=true,Style*=--~)
& \textbf{Recurrent workers}. While returning workers are desirable in any crowdsourcing task, they represent a potential source of bias in the context of task design evaluation, i.e.,  
they might perform better, due to the \textit{learning} effect.
In our experiments, we observed a 38\% of returning workers, who featured a lower completion time but not higher accuracy\footnote{We noticed, however, that accuracy remained mostly unaffected by conditions and other factors across all our experiments, and it might have been less susceptible to the learning effect.} (Figure \ref{fig:flow}A).


& \textbf{Condition crossover}. Returning workers can also land in a different experimental condition,
which could modify their behavior and performance. From the 30\% workers who crossed conditions (Figure \ref{fig:flow}A), 
we observed that switching between highlighting support and not support resulted in lower decision time, but that having had a bad highlighting support before can increase the time when dealing with good support - possibly due to the lack of trust in the support. Workers switching from support to no support also featured higher accuracy than the new workers and those returning to the same condition.




& \textbf{Timezones}. Running the conditions at different times can introduce confounding factors that could hurt the comparison. For instance, we observed the worker performance in independent runs varied by different factors even between runs of the same condition 
(e.g., from 24s to 14s in decision time between a first and a second run considering only new workers)
Thus collecting reliable and comparable results in this setting would require multiple runs over a long period of time.

& \textbf{Population demographics}. The pool of active workers is determined by the demographics of the crowdsourcing platform and the time an experiment runs.
In running uncontrolled tasks, we observed a participation dominated by certain countries, which prevented 
more diverse
population characteristics. For example, the top contributing countries provided 48.1\% of the total judgements 
(Venezuela: 28.5\%, Egypt: 11.8\%, Ukraine: 7.8\%).




\end{easylist}

Running a systematic comparison of task designs using the native building blocks of a crowdsourcing platform is thus a complex activity, susceptible to different types of experimental biases, which are costly to clean up (e.g., discarding 38\% of the dataset)
%
- a challenge many  task designers and researchers face.

\section{CrowdHub Platform}


The above challenges motivated us to design and build a tool that extends the capabilities of crowdsourcing platforms for the purpose of task evaluation. CrowdHub is a system that sits on top of major crowd platforms, such as F8 and Toloka, and offers the building blocks to design and run controlled experiments using crowdsourcing.






We based our design on the following main ideas: workflows, eligibility control, population management and time sampling.
\textit{\textbf{Workflows}} allow requesters to set the foundation for their experimental designs by defining the tasks and sequence of execution (sequential or parallel) --- reducing issues in working with different active crowds. 
%
The \textit{\textbf{eligibility control}} of tasks allows requesters to define the policy regarding returning workers and condition crossovers associated with the experimental design (between- or within-groups design).
%
Through \textit{\textbf{population management}} requesters can control for subgroups of workers dominating a dataset by assigning a specific quota, and \textit{\textbf{time sampling}} helps in controlling for confounding factors by scheduling task execution over a period of time.
Altogether, these features allow requester to be in control of their experimental design. 




CrowdHub enables the entire task evaluation process, as shown in Figure \ref{fig:flow}. At \textit{design time}, requesters use the workflow editor to define the experimental design, which includes the tasks (\textit{Do} boxes) and the data flow (indicated by the arrows and the \textit{lambda} functions describing data aggregation and partitioning). Experimental groups can also be defined and associated to one or more tasks, denoted in the diagram using different colors. When deploying the experiment, CrowdHub parses the workflow definition and creates the individual tasks in the target crowdsourcing platform with the associated data units and task design. At \textit{run time}, the requestor can specify the population management strategy, and time sampling, if any, and the platform will launch, pause and resume the tasks, and constraint the workers access to tasks, accordingly. 
From a technical perspective, CrowdHub manages the interactions with the crowdsourcing platform through their public APIs and JavaScript extensions incorporated to the tasks, for additional metrics, worker control, and worker identification (browser fingerprinting) \cite{gadiraju2017improving}.

In this demo we will show the entire evaluation lifecycle with the current version of CrowdHub \footnote{\url{https://github.com/TrentoCrowdAI/crowdhub-web}}, including workflow design, eligibility control and deployment on F8 and Toloka. 

\bibliographystyle{aaai}
\bibliography{references}

\begin{thebibliography}{}

\bibitem[\protect\citeauthoryear{Difallah \bgroup et al\mbox.\egroup
  }{2014}]{difallah2014scaling}
Difallah, D.~E.; Catasta, M.; Demartini, G.; and Cudr{\'e}-Mauroux, P.
\newblock 2014.
\newblock Scaling-up the crowd: Micro-task pricing schemes for worker retention
  and latency improvement.
\newblock In {\em Second AAAI Conference on Human Computation and
  Crowdsourcing}.

\bibitem[\protect\citeauthoryear{Drapeau \bgroup et al\mbox.\egroup
  }{2016}]{drapeau2016microtalk}
Drapeau, R.; Chilton, L.~B.; Bragg, J.; and Weld, D.~S.
\newblock 2016.
\newblock Microtalk: Using argumentation to improve crowdsourcing accuracy.
\newblock In {\em Fourth AAAI Conference on Human Computation and
  Crowdsourcing}.

\bibitem[\protect\citeauthoryear{Eickhoff}{2018}]{eickhoff2018cognitive}
Eickhoff, C.
\newblock 2018.
\newblock Cognitive biases in crowdsourcing.
\newblock In {\em Proceedings of the Eleventh ACM International Conference on
  Web Search and Data Mining},  162--170.
\newblock ACM.

\bibitem[\protect\citeauthoryear{Faltings \bgroup et al\mbox.\egroup
  }{2014}]{DBLP:conf/hcomp/FaltingsJPT14}
Faltings, B.; Jurca, R.; Pu, P.; and Tran, B.~D.
\newblock 2014.
\newblock Incentives to counter bias in human computation.
\newblock In {\em Proceedings of the Seconf {AAAI} Conference on Human
  Computation and Crowdsourcing, {HCOMP} 2014, November 2-4, 2014, Pittsburgh,
  Pennsylvania, {USA}}.

\bibitem[\protect\citeauthoryear{Gadiraju and
  Kawase}{2017}]{gadiraju2017improving}
Gadiraju, U., and Kawase, R.
\newblock 2017.
\newblock Improving reliability of crowdsourced results by detecting crowd
  workers with multiple identities.
\newblock In {\em International Conference on Web Engineering},  190--205.
\newblock Springer.

\bibitem[\protect\citeauthoryear{Gadiraju \bgroup et al\mbox.\egroup
  }{2017}]{DBLP:journals/imwut/GadirajuCGD17}
Gadiraju, U.; Checco, A.; Gupta, N.; and Demartini, G.
\newblock 2017.
\newblock Modus operandi of crowd workers: The invisible role of microtask work
  environments.
\newblock {\em {IMWUT}} 1(3):49:1--49:29.

\bibitem[\protect\citeauthoryear{Gadiraju, Yang, and
  Bozzon}{2017}]{DBLP:conf/ht/GadirajuYB17}
Gadiraju, U.; Yang, J.; and Bozzon, A.
\newblock 2017.
\newblock Clarity is a worthwhile quality: On the role of task clarity in
  microtask crowdsourcing.
\newblock In {\em Proceedings of the 28th {ACM} Conference on Hypertext and
  Social Media, {HT} 2017, Prague, Czech Republic, July 4-7, 2017},  5--14.

\bibitem[\protect\citeauthoryear{Jain \bgroup et al\mbox.\egroup
  }{2017}]{DBLP:journals/pvldb/JainSPW17}
Jain, A.; Sarma, A.~D.; Parameswaran, A.~G.; and Widom, J.
\newblock 2017.
\newblock Understanding workers, developing effective tasks, and enhancing
  marketplace dynamics: {A} study of a large crowdsourcing marketplace.
\newblock {\em {PVLDB}} 10(7):829--840.

\bibitem[\protect\citeauthoryear{Kamar, Kapoor, and
  Horvitz}{2015}]{DBLP:conf/hcomp/KamarKH15}
Kamar, E.; Kapoor, A.; and Horvitz, E.
\newblock 2015.
\newblock Identifying and accounting for task-dependent bias in crowdsourcing.
\newblock In {\em Proceedings of the Third {AAAI} Conference on Human
  Computation and Crowdsourcing, {HCOMP} 2015, November 8-11, 2015, San Diego,
  California, {USA.}},  92--101.

\bibitem[\protect\citeauthoryear{Maddalena \bgroup et al\mbox.\egroup
  }{2016}]{maddalena2016crowdsourcing}
Maddalena, E.; Basaldella, M.; De~Nart, D.; Degl'Innocenti, D.; Mizzaro, S.;
  and Demartini, G.
\newblock 2016.
\newblock Crowdsourcing relevance assessments: The unexpected benefits of
  limiting the time to judge.
\newblock In {\em Fourth AAAI Conference on Human Computation and
  Crowdsourcing}.

\bibitem[\protect\citeauthoryear{Ram\'{i}rez \bgroup et al\mbox.\egroup
  }{2019}]{ramirez2019}
Ram\'{i}rez, J.; Baez, M.; Casati, F.; and Benatallah, B.
\newblock 2019.
\newblock Understanding the impact of text highlighting in crowdsourcing tasks.
\newblock In {\em Proceedings of the Seventh {AAAI} Conference on Human
  Computation and Crowdsourcing, {HCOMP} 2019}.

\bibitem[\protect\citeauthoryear{Sampath, Rajeshuni, and
  Indurkhya}{2014}]{DBLP:conf/chi/SampathRI14}
Sampath, H.~A.; Rajeshuni, R.; and Indurkhya, B.
\newblock 2014.
\newblock Cognitively inspired task design to improve user performance on
  crowdsourcing platforms.
\newblock In {\em {CHI} Conference on Human Factors in Computing Systems,
  CHI'14, Toronto, ON, Canada - April 26 - May 01, 2014},  3665--3674.

\end{thebibliography}

\end{document}